\documentclass[preprint,floats,aps,12pt,superscriptaddress,nofootinbib,floatfix]{revtex4}
\usepackage{natbib} 
\usepackage{amssymb,amsmath}
\usepackage{csquotes}
\usepackage[usenames,dvipsnames]{color}
\usepackage[pdftex]{graphicx}
\usepackage[pdftex]{hyperref}
\hypersetup{pdftitle={Gradient dynamics model for drops spreading on polymer brushes}, pdfauthor={Uwe Thiele and Simon Hartmann},  pdfproducer={lateX},
pdfview=FitV,       pdfstartview=FitB,
linkcolor=blue,     citecolor=blue,     urlcolor=red,      breaklinks=true,    colorlinks=true,
citebordercolor=0 0 0,  filebordercolor=0 0 0,
linkbordercolor=0 0 0,
menubordercolor=0 0 0,
urlbordercolor=0 0 0,
pdfhighlight=/I,
pdfborder=0 0 0,   bookmarksopen=true,
bookmarksnumbered=true
}
 \graphicspath{{.},{./Figures/}} \renewcommand{\vec}[1]{\mathbf{#1}}

\newcommand{\tens}[1]{\mathbf{\underline{#1}}}

\begin{document}
\title{Gradient dynamics model for drops spreading on polymer brushes}
\author{Uwe Thiele}
 \email{u.thiele@uni-muenster.de}
 \homepage{http://www.uwethiele.de}
 \affiliation{Institut f\"ur Theoretische Physik, Westf\"alische Wilhelms-Universit\"at M\"unster, Wilhelm Klemm Str.\ 9, 48149 M\"unster, Germany}
 \affiliation{Center of Nonlinear Science (CeNoS), Westf{\"a}lische Wilhelms-Universit\"at M\"unster, Corrensstr.\ 2, 48149 M\"unster, Germany}
 \affiliation{Center for Multiscale Theory and Computation (CMTC), Westf{\"a}lische Wilhelms-Universit\"at, Corrensstr.\ 40, 48149 M\"unster, Germany}
\author{Simon Hartmann}
 \email{s.hartmann@uni-muenster.de}
\affiliation{Institut f\"ur Theoretische Physik, Westf\"alische Wilhelms-Universit\"at M\"unster, Wilhelm Klemm Str.\ 9, 48149 M\"unster, Germany}

\begin{abstract} 
When a liquid drop spreads on an adaptive substrate the latter changes its properties what may result in an intricate coupled dynamics of drop and substrate. Here we present a generic mesoscale hydrodynamic model for such processes that is written as a gradient dynamics on an underlying energy functional. We specify the model details for the example of a drop spreading on a dry polymer brush. There, liquid absorption into the brush results in swelling of the brush causing changes in the brush topography and wettability. The liquid may also advance within the brush via diffusion (or wicking) resulting in coupled drop and brush dynamics. The specific model accounts for coupled spreading, absorption and wicking dynamics when the underlying energy functional incorporates capillarity, wettability and brush energy. After employing a simple version of such a model to numerically simulate a droplet spreading on a swelling brush we conclude with a discussion of possible model extensions.
\end{abstract}
\maketitle
\section{Introduction}\label{sec:intro}
In spreading processes simple or complex liquids advance onto various substrates. Such dynamic wetting processes are common in daily life and are also of large importance for many technological processes~\cite{GennesBrochard-WyartQuere2004,KalliadasisThiele2007,StarovVelardeRadke2007,Bormashenko2017}. Most experimental and theoretical work of the past decades considers these processes on smooth homogeneous solid substrates or studies the influence of static substrate heterogeneities like wettability and topography patterns and defects~\cite{TBBB2003epje,HeBS2008armr,BEIM2009rmp,MiPi2017ap}.
However, recent developments in areas like microelectronics or 3D printing increasingly involve cases where (de)wet\-ting hydrodynamics and substrate dynamics are coupled. This is particularly important on microscopic and mesoscopic length scales, where (non-)equilibrium interface phenomena dominate. For instance, viscous and soft elastic substrates reversibly change their profile when one deposits a liquid drop~\cite{BCJP2013epje,LWBD2014jfm,PAKL2009sm,StDu2012sm,AnSn2020arfm}. In this case, nearly no transport of material takes place across the liquid-solid interface and the substrate mainly changes its topography. 

In contrast, adaptive substrates change their physico-chemical properties like wettability and possibly additionally their topography due to the presence of a liquid or through external conditions~\cite{GuGu2016ra,CHGM2010nm}. This may be induced by direct contact like under a sessile droplet, or it may be mediated through an external phase like the ambient atmosphere or a second liquid. Further, modifications may be mediated through gradual changes of temperature~\cite{MMTM2016an,CLLZ2010sm,SRLB2017aami,PLMB2016mme}, ambient humidity~\cite{KaLR2017l} and pH-value~\cite{CHGM2010nm,GuGu2016ra,YWJS2005am,ZhZW2012nam,SZZF2012jcis,CLJZ2014am}. A recent overview of experimental systems and theoretical approaches is given in Ref.~\cite{BBSV2018l}.

For adaptive substrates, transport of material may take place across the liquid-solid interface and phase boundaries may shift. For instance, polymer brushes and networks, hydrogels and organic multilayers may swell under a drop of liquid or in response to its extending vapour~\cite{HaFK2007sm,CHGM2010nm}. In a co-nonsolvency transition poly(N-isopropylacrylamide) (PNiPAAm) brushes in aqueous ethanol mixtures show a phase transition including a nonmonotonic dependency of brush thickness on solvent concentration~\cite{YREU2018m}. Similar effects are observed for hydrogel substrates~\cite{BKTW2017aml,BKTW2017l}.
In all cases, the dynamics of the adaptive substrate and the dynamics of the de(wetting) liquid take place on similar time scales, i.e., their direct coupling may results in new phenomena. A limiting case of ultra-thin adaptive surface layers are the mentioned substrates covered with a polymer brush~\cite{PLMB2016mme,TGMK2012am,MMMN2003jacs}. For very mobile brushes one can create liquid-like surfaces characterised by very small lateral adhesion forces. A common model system  are PDMS-brushes. The mobility results in a relatively fast change of the brush thickness in the three-phase contact line region, i.e., the effective substrate profile is adaptive. One also expects that the contact line motion orders the brush molecules resulting in further anisotropy effects. Therefore the dynamic wetting behaviour is influenced by the speed of adaptation, i.e.\ results from the interplay of intermolecular forces of the coating, the vapour pressure of the liquid and the adhesion forces.

The theoretical description and numerical simulation of wetting processes on micro-, meso- and macroscales is in the case of simple liquids on inert solid substrates quite well developed. The range of approaches includes Molecular Dynamics (MD) simulations~\cite{DeBl2008armr,LSPM2011jpm,DRKH2012sm}, lattice Boltzmann simulations~\cite{DRKH2012sm}, phase-field models~\cite{YuFe2011epjt}, classical hydrodynamics (Navier-Stokes equations)~\cite{MADK2013pf,MaAK2016pf} and asymptotic approaches as mesoscopic thin-film (or long-wave) models~\cite{OrDB1997rmp,CrMa2009rmp}. Comparative studies, parameter passing approaches and consistency conditions connect these approaches into a multiscale framework~\cite{DRKH2012sm,MADK2013pf,TMTT2013jcp,SNSK2015jem,BoTH2018jfm}. In macroscopic hydrodynamic models, polymer brushes are often considered as flexible (viscoelastic) layers without adaptability~\cite{LoAL1996l}. Ref.~\cite{BBSV2018l} presents a macroscopic dynamical model that incorporates the effect of an adaptive substrate based on an imposed exponentially relaxing equilibrium contact angle. The interaction of liquid drops with polymer brushes is also studied employing MD simulations \cite{LeMu2011jcp,MeSB2019m}. Here, we shall pursue a continuum model for the coupled drop and brush dynamics.

A versatile asymptotic method to study nonequilibrium thin films and shallow drops are thin-film models derived via an long-wave expansion from the governing equations and boundary conditions of hydrodynamics~\cite{OrDB1997rmp,Thie2007chapter}. They can often be brought into gradient dynamics form~\cite{Mitl1993jcis,EGUW2019springer}. Then, for a drop or film of simple nonvolatile liquid on a flat solid substrate the mass-conserving dynamics for the film height profile $h(x,y,t)$ is written as
\begin{equation}
\partial_t h \,=\,
\nabla\cdot\left[Q(h)\nabla\frac{\delta \mathcal{F}}{\delta h}\right]
\quad\mathrm{with}\quad
\mathcal{F}[h]\,=\,\int\left[\frac{\gamma}{2}{(\nabla h)}^2 + f(h) \right]d^2r
\label{eq:onefield:gov}
\end{equation}
where the energy functional $\mathcal{F}[h]$ accounts for capillarity and wettability, $\gamma$ is the liquid-gas interfacial energy, $f(h)$ a film height-dependent wetting energy (or binding potential)~\cite{Chur1995acis,Genn1985rmp,TeDS1988rpap,StVe2009jpm}, $Q(h)=h^3/3\eta$ is the mobility function in the case without slip at the substrate, and $\eta$ is the dynamic viscosity. 
In general, different descriptions of a moving contact line are still under debate (see, e.g.,~\cite{Vela2011epjst} and other contributions in the corresponding discussion volume). In the context of thin-film models the two major approaches are slip models where it is assumed that the liquid slips at the solid substrate (amending the mobility $Q$~\cite{MuWW2005jem}) and precursor film models that assume an ultrathin adsorption layer exists on the macroscopically dry substrate~\cite{PODC2012jpm} as determined by the wetting energy.

For selected complex liquids, a theoretical description of the interaction of (de)wetting dynamics and the dynamics of the internal degrees of freedom of the liquid also exists. For instance, mesoscopic thin-film models describe films and drops of mixtures of simple liquids and surfactant solutions on homogeneous, solid and inert substrates~\cite{MaCr2009sm,NaTh2010n,KaRi2014jfm}. Such models can often be reformulated as gradient dynamics on an underlying energy functional~\cite{Thie2018csa}. This then allows for systematic and fully thermodynamically consistent extensions~\cite{ThTL2013prl,ThAP2016prf} that are able to capture the full extent of the interface-dominated dynamics in cases where the various diffusive and advective transport channels couple, e.g., in the case of surfactant-dependent wettability~\cite{TSTJ2018l}. Further, the gradient dynamics approach provides a simple criterion to assess the validity of other models. 

To our knowledge the coupled dynamics of an adaptive substrate and a spreading liquid drop has not yet been considered with a mesoscopic hydrodynamic model. It is the aim of the present work to develop such a mesoscopic hydrodynamic model using as an example a drop of simple liquids spreading on a simple polymer brush. The model is written as a gradient dynamics on an underlying energy functional that accounts for capillarity, wettability and brush energy. For the latter we use the Alexander-de Gennes approach~\cite{Genn1991casi,Alex1977jp,Somm2017m}. This shall in the future allow for a number of extensions towards more complicated brush behaviour.

Our work is structured as follows. In section~\ref{sec:model} we introduce the dynamical model and the underlying energy. The subsequent section~\ref{sec:res} presents selected numerical results for droplets spreading on swelling brushes and compares our results with other approaches in the literature. Finally, section~\ref{sec:conc} concludes with a discussion of possible future model extensions.

\section{Dynamic model for simple liquid on adaptive substrate}\label{sec:model}
\subsection{Gradient dynamics}
\begin{figure}[tbh]
  \centering
  \includegraphics[width=0.9\hsize]{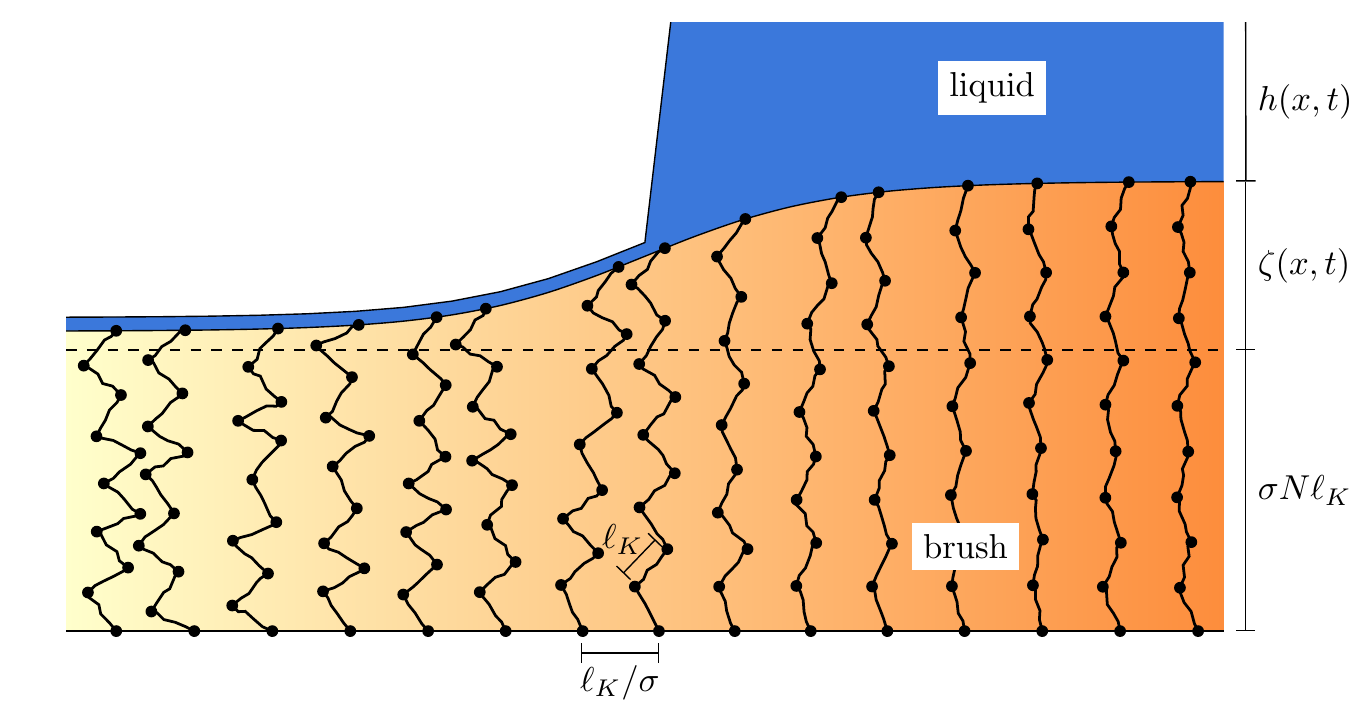}
  \caption{Sketch of the brush at a three-phase contact line region with a liquid film of height $h(x,t)$. The relative grafting density $\sigma$ and the chain length $N\ell_K$ determine the dry (collapsed) brush height. Liquid absorption leads to expansion of the polymers and an increase of the brush height by $\zeta(x,t)$, i.e., the effective height of the absorbed liquid, indicated by the color gradient.}\label{fig:buerste}
\end{figure}

Here, we develop a generic model for thin films/shallow drops of simple liquid on adaptive substrates. We consider relatively simple substrates that can be characterised by a single order parameter field that is then related to, e.g., the local stretching state of a polymer brush covering the bare substrate (see sketch Fig.~\ref{fig:buerste}). In particular, we use a generic quantity \enquote{substrate-absorbed liquid} as order parameter field. It represents the liquid absorbed into the adapting surface layer of the substrate. It can, e.g., be directly related to the stretching state of a grafted polymer brush or to the filling ratio of a porous layer. This absorption is a proper thermodynamic quantity, i.e., at equilibrium it is determined by an appropriate energy functional. The gradient dynamics approach then allows for a description of the coupled dynamics of the thickness profile $h(\mathrm{x},t)$ of the liquid above the substrate and of the profile of substrate-absorbed liquid $\zeta(\mathrm{x},t)$. Both represent (effective) heights and have units of length. The local thickness of the brush layer $H$ is directly related to $\zeta$ (see below section~\ref{sec:en-brush}). The general gradient dynamics of two scalar fields has the form (see equation~(4) of~\cite{Thie2018csa} with $n=2$)
\begin{align}
\partial_t h \,&=\,
\nabla\cdot\left[Q_{hh} \nabla\frac{\delta F}{\delta h} + Q_{h\zeta} \nabla\frac{\delta F}{\delta \zeta}\right] 
- M_{hh} \frac{\delta F}{\delta h} - M_{h\zeta} \frac{\delta F}{\delta \zeta}
\label{eq:twofield-evolution-one}\\
\partial_t \zeta \,&=\,
\nabla\cdot\left[Q_{\zeta h} \nabla\frac{\delta F}{\delta h} + Q_{\zeta \zeta} \nabla\frac{\delta  F}{\delta \zeta}\right]
- M_{\zeta h} \frac{\delta F}{\delta h} - M_{\zeta \zeta} \frac{\delta  F}{\delta \zeta}
\label{eq:twofield-evolution-two}
\end{align}
Note that in the general case, the dynamics of each of the fields combines a mass-conserving part (the respective first r.h.s.\ term, of the form of a divergence of a flux $\nabla\cdot\vec{j}$) and a nonmass-conserving part (the respective second r.h.s.\ term, a transfer rate $r$) that are both governed by the same underlying energy functional $F[h,\zeta]$. We call the terms \enquote{conserved} and 
\enquote{nonconserved}, respectively. The corresponding mobility functions form the positive definite and symmetric $2\times2$ matrices $\tens{Q}$ and $\tens{M}$. Examples and more background for two-field systems and a three-field example are given in~\cite{ThAP2016prf}. The system (\ref{eq:twofield-evolution-one})--(\ref{eq:twofield-evolution-two}) is generic and can be adapted for different adaptive substrates. Also the consideration concerning the mobilities in the remainder of the section is generic and one may replace \enquote{brush} by \enquote{adaptive substrate}. However, to be specific the argument is developed for the case of a polymeric brush pictured in Fig.~\ref{fig:buerste}.

For simplicity, here, we assume that the dynamics encompasses only three processes: (i) the hydrodynamic motion of the liquid in the film/drop, (ii) the local transfer of liquid between the film and the brush, and (iii) the diffusion of liquid within the brush. This implies that there is no dynamic coupling of hydrodynamic motion within the drop/film and the brush, i.e., $Q_{h\zeta}=Q_{\zeta h}=0$. Also neglecting slip at the brush-liquid boundary, the conserved mobilities are 
\begin{equation}
\mathbf{Q}
=\left(\!\!\!
\begin{array}{cc}  
h^3/3 \eta & 0 \\[.3ex]
0& D \zeta
\end{array}
\!\!\!\right)
\nonumber \label{eq:twofield-mob-c}
\end{equation}
where $D$ is a diffusive mobility. Without the nonconserved terms and energetic coupling, i.e., $F=F[h]$, Eq.~(\ref{eq:twofield-evolution-one}) reduces to the standard thin-film equation (\ref{eq:onefield:gov}). Similarly, for $F=F[\zeta]$, the decoupled Eq.~(\ref{eq:twofield-evolution-two}) becomes a (possibly non-Fickian) diffusion equation describing \enquote{wicking} within a polymer brush.

Without evaporation, $h+\zeta$ is conserved, i.e.,  $\partial_t (h+\zeta)$ equals the divergence of the total flux. Therefore, the simplest nonconserved mobilities are
\begin{equation}
\mathbf{M}
= M\left(\!\!\!
\begin{array}{cc}  
1 & -1 \\[.3ex]
-1& 1
\end{array}
\!\!\!\right)
\nonumber \label{eq:twofield-mob-nc}
\end{equation}
where $M$ is a transfer rate constant. We argue that Eqs.~(\ref{eq:twofield-evolution-one})--(\ref{eq:twofield-mob-nc}) represent the simplest thermodynamically consistent framework model for the spreading of a drop on a polymer brush. It accounts for the dynamic processes of liquid convection via $Q_{hh}$, diffusion in the brush via $Q_{\zeta \zeta}$, and transfer between drop and brush via $M$. All dynamic processes are consistently driven by the same energy functional $F[h,\zeta]$ that we discuss next. Note, that the form of Eqs.~(\ref{eq:twofield-evolution-one})--(\ref{eq:twofield-mob-nc}) may be applied for other adaptive substrates, where adaptivity is related to mass transfer. 
\subsection{Energy functional for drop on polymeric brush}\label{sec:en-brush}
For sufficiently small drops or sufficiently thin films of a simple liquid on a brush we can neglect gravity and the governing energy functional only needs to capture capillarity, i.e., interfacial energies, wettability and the energy determining the state of the brush. We write
\begin{equation}
F[h,\zeta] = \int \left[
f_\mathrm{cap}(h,\zeta) + f_\mathrm{wet}(h,\zeta) + g_\mathrm{brush}(\zeta)
\right] d^2x.
\label{eq:brush-en-functional}
\end{equation}
The capillarity contribution
\begin{equation}
f_\mathrm{cap}(h,\zeta) = \frac{\gamma}{2}|\nabla (h+\zeta)|^2 + \frac{\gamma_\mathrm{bl}}{2}|\nabla \zeta|^2
\label{eq:brush-en-cap}
\end{equation}
contains the energy of the liquid-gas and the brush-liquid interface in long-wave approximation (with interfacial tensions $\gamma$ and $\gamma_\mathrm{bl}$, here, both taken constant). For the wettability contribution we use the simple ansatz for partially wetting liquids
\begin{equation}
f_\mathrm{wet}(h, \zeta)=f_\mathrm{wet}(h) = -\frac{A}{2h^2} + \frac{B}{5h^5} 
\label{eq:brush-en-wet}
\end{equation}
where $A$ and $B$ are Hamaker-type constants for long- and short-range contributions, respectively. Here, for simplicity, we explicitly exclude a dependency of wettability and brush-liquid interface tension on brush state (see discussion in section~\ref{sec:conc}).

Finally, for the brush energy we adapt the Alexander-de Gennes mean-field approach~\cite{Alex1977jp,Genn1991casi}. Specifically, we employ the form given in~\cite{Somm2017m} for a brush interacting with a binary mixture [their Eq.~(1c)] setting their co-nonsolvent concentration $\phi$ to zero. The resulting energy per monomer unit of the brush is
\begin{equation}
  \hat g_\mathrm{brush} = \frac{k_B T}{2} \frac{\sigma^2}{c^2} + k_B T \left(
  \frac{1}{c} -1\right)\,\log(1-c),
\label{eq:brush-en-mono}
\end{equation}
where $c$ is the volume fraction of brush monomers within the brush layer and $\sigma$ is the relative grafting density, i.e., an area fraction.  Eq.~(\ref{eq:brush-en-mono}) combines the conformational free energy of the brush and the mixing free energy of (miscible) brush and solvent.  The term $\sim \sigma^2/c^2$ is the stretching energy in Gaussian approximation while the entire second term is the Flory-Huggins free energy per brush monomer. Note, that the relation between brush volume fraction $c$ and the brush layer thickness is $H = \sigma N \ell_K / c$ with the Kuhn segment length $\ell_K$. The relation arises from equating two expressions for the total polymer volume per substrate area $A$, namely, as $\sigma N \ell_K A$ (completely dry brush, i.e., fully compacted) or as $c H A$.
Assuming polymer molecules do not overlap, $\sigma$ is proportional to the dry brush height $H(\zeta = 0) = \sigma N \ell_K$. The brush layer thickness depends on absorption as $H = \zeta + \sigma N \ell_K$. The chain length $N$ does not explicitly enter the free energy per brush monomer (\ref{eq:brush-en-mono}), i.e., the phase behaviour of brushes with identical grafting density is identical as long as brush molecules do not overlap in the collapsed state~\cite{Somm2017m}.

Next, the brush energy per monomer $\hat g_\mathrm{brush}$ is transformed into an absorption-dependent brush energy per substrate area $g_\mathrm{brush}=N\sigma_\mathrm{abs} \hat g_\mathrm{brush}$ where $\sigma_\mathrm{abs} = \sigma /\ell_K^2$ is the absolute grafting density.  This energy shall be expressed in terms of solvent absorption into the brush $\zeta=(1-c) H = (1-c)\sigma N \ell_K / c$.  The latter implies we can express $c=\sigma N \ell_K / (\zeta + \sigma N \ell_K)$. Bringing everything together, we have the brush energy
\begin{equation}
g_\mathrm{brush}(\zeta) = \frac{\sigma N \ell_K k_B T}{\ell_K^3}\left[\frac{\sigma^2}{2} {\left(\frac{\zeta + \sigma N \ell_K}{\sigma N \ell_K}\right)}^2 +
\frac{\zeta}{\sigma N \ell_K} \,\log\left(\frac{\zeta}{\zeta + \sigma N \ell_K}\right) \right].
\label{eq:brush-en-area}
\end{equation}
Next, we nondimensionalize the full set of equations.
\subsection{Nondimensionalization}\label{sec:nd-brush}
The evaluation of the energy functional variations gives
\begin{equation}
    \begin{aligned}
        \frac{\delta F}{\delta h} &= -\gamma \Delta (h + \zeta) + \frac{A}{h^3} - \frac{B}{h^6}\\
        \frac{\delta F}{\delta \zeta} &= -\gamma \Delta (h + \zeta) - \gamma_\mathrm{bl} \Delta \zeta + g'_\mathrm{brush}(\zeta)
    \end{aligned}
\end{equation}
with the derivative of the local brush energy
\begin{equation}
    g'_\mathrm{brush}(\zeta) = \frac{k_B T}{\ell_K^3} \left[ \sigma^2\, \frac{\zeta + \sigma N \ell_K}{\sigma N \ell_K} + \frac{\sigma N \ell_K}{\zeta + \sigma N \ell_K} + \log\left(\frac{\zeta}{\zeta + \sigma N \ell_K}\right) \right].
\end{equation}

We further express the short-range interaction strength $B$ by using the explicit height of the precursor layer $h_p = \sqrt[3]{B/A}$. Since the wetting potential $f_\mathrm{wet}$ determines the equilibrium contact angle as $\cos \theta_e = 1 + \frac{1}{\gamma} f_\mathrm{wet}(h_p)$~\cite{Chur1995acis}, for small contact angles the remaining Hamaker constant may be substituted by $A=\frac{5}{3}\gamma h_p^2 \theta_e^2$.

We introduce length scales $h_0=\zeta_0=h_p$ and $x_0=\sqrt{3/5}\, h_p / \theta_e$, a time scale $t_0 = 27 \eta h_p/(25 \gamma \theta_e^4)$ and an energy (per area) scale $F_0 = 5 \gamma \theta_e^2/3$ in order to rewrite the model into a dimensionless form. The free energy variations then become
\begin{equation}
    \begin{aligned}
        \frac{\delta F}{\delta h} &= -\Delta (h + \zeta) + \frac{1}{h^3} - \frac{1}{h^6}\\
        \frac{\delta F}{\delta \zeta} &= -\Delta (h + \zeta) - \tilde \gamma_\mathrm{bl} \Delta \zeta + \frac{\sigma \tilde T}{\tilde l} \left(\zeta + \sigma \tilde l \right) + \tilde T \left[ \frac{\sigma \tilde l}{\zeta + \sigma \tilde l} + \log\left(\frac{\zeta}{\zeta + \sigma \tilde l}\right) \right]
    \end{aligned}
\end{equation}
where the remaining dimensionless parameters are
\begin{equation}
    \tilde \gamma_\mathrm{bl} = \frac{\gamma_\mathrm{bl}}{\gamma},~~~~~~
    \tilde l = \frac{N\ell_K}{h_p},~~~~~~
    \tilde T = \frac{3 h_p k_B T}{5 \gamma \theta_e^2 \ell_K^3}.
\end{equation}
Similarly, the gradient dynamics reduce to
\begin{equation}
    \begin{aligned}
        \partial_t h &= \nabla \cdot \left[ \,h^3\hspace{0.2em} \nabla \frac{\delta F}{\delta h}\right] - \tilde M \left[\frac{\delta F}{\delta h} - \frac{\delta F}{\delta \zeta}\right]\\
        \partial_t \zeta &= \nabla \cdot \left[ \tilde D \zeta \nabla \frac{\delta F}{\delta \zeta}\right] + \tilde M \left[\frac{\delta F}{\delta h} - \frac{\delta F}{\delta \zeta}\right]\label{eq:twofield-evolution-nondim}
    \end{aligned}
\end{equation}
where we use
\begin{equation}
    \tilde D = \frac{3\eta D}{h_p^2},~~~~~~~~~~~
    \tilde M = 3\eta \gamma \frac{3 M}{5\gamma h_p \theta_e^2}
\end{equation}
as dimensionless constants for diffusion and transfer rates.

The above procedure effectively reduces the complexity of the parameter space and scales the dynamics for optimal numerical treatment. In the following, we employ a finite element approach using the C++ FEM library \textsc{oomph-lib}~\cite{HeHa2006} for time simulations of the governing equations in one spatial dimension. For time-stepping, a backward differentiation formula (BDF) scheme of second order is used, while both the time-steps and the spatial mesh are adaptive to the dynamics.

\section{Results --- drop spreading on a brush}\label{sec:res}
\begin{figure}[tbh]
    \centering
    \includegraphics[width=\hsize]{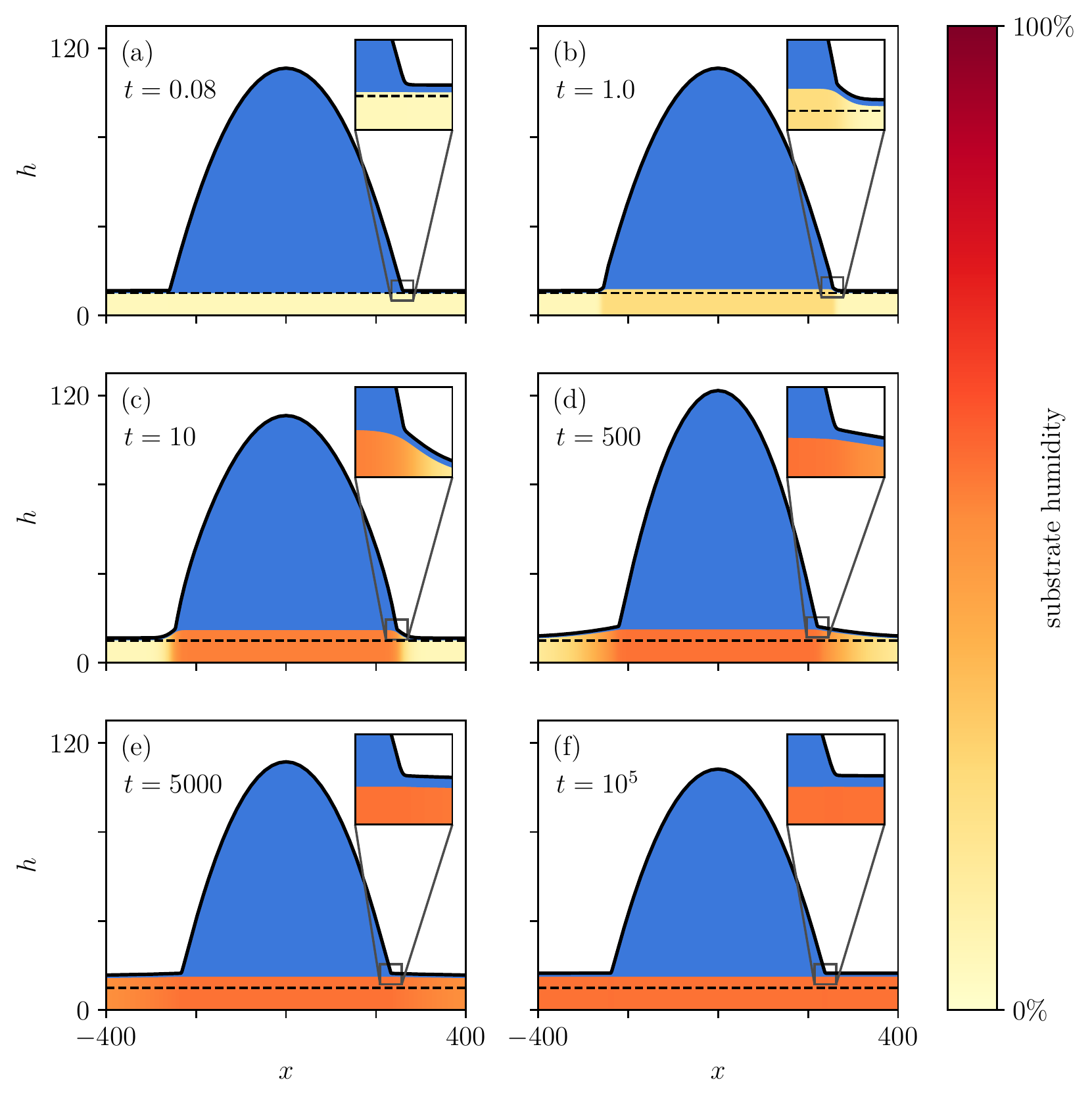}
    \caption{
        The snapshots show the states of a spreading liquid drop on an initially dry polymer brush at times (a) $t=0.08$,  (b) $t=1.0$,  (c) $t=10.0$,  (d) $t=500.0$,  (e) $t=5\cdot 10^3$ and  (f) $t=10^5$. The height profiles of the liquid drop $h+H$ and of the brush $H$ are shown as solid lines while the absorbed amount of liquid in the brush $\zeta$ is indicated as humidity $\zeta/H_\mathrm{max}$ (with $H_\mathrm{max}=(1-\sigma)N\ell_K$) by a yellow to orange shading of the brush layer (see colour bar). The horizontal dashed line indicates the reference height of a completely dry brush. The insets show the respective contact line regions and the dimensionless parameters are $\sigma=0.5$, $\tilde l = 20$, $\tilde \gamma_\mathrm{bl} = 50$, $\tilde T=22$, $\tilde D=1$ and $\tilde M=0.1$.}\label{fig:buerste-spreading-snapshots}
\end{figure}

\begin{figure}[tbh]
  \centering
  \includegraphics[width=\hsize]{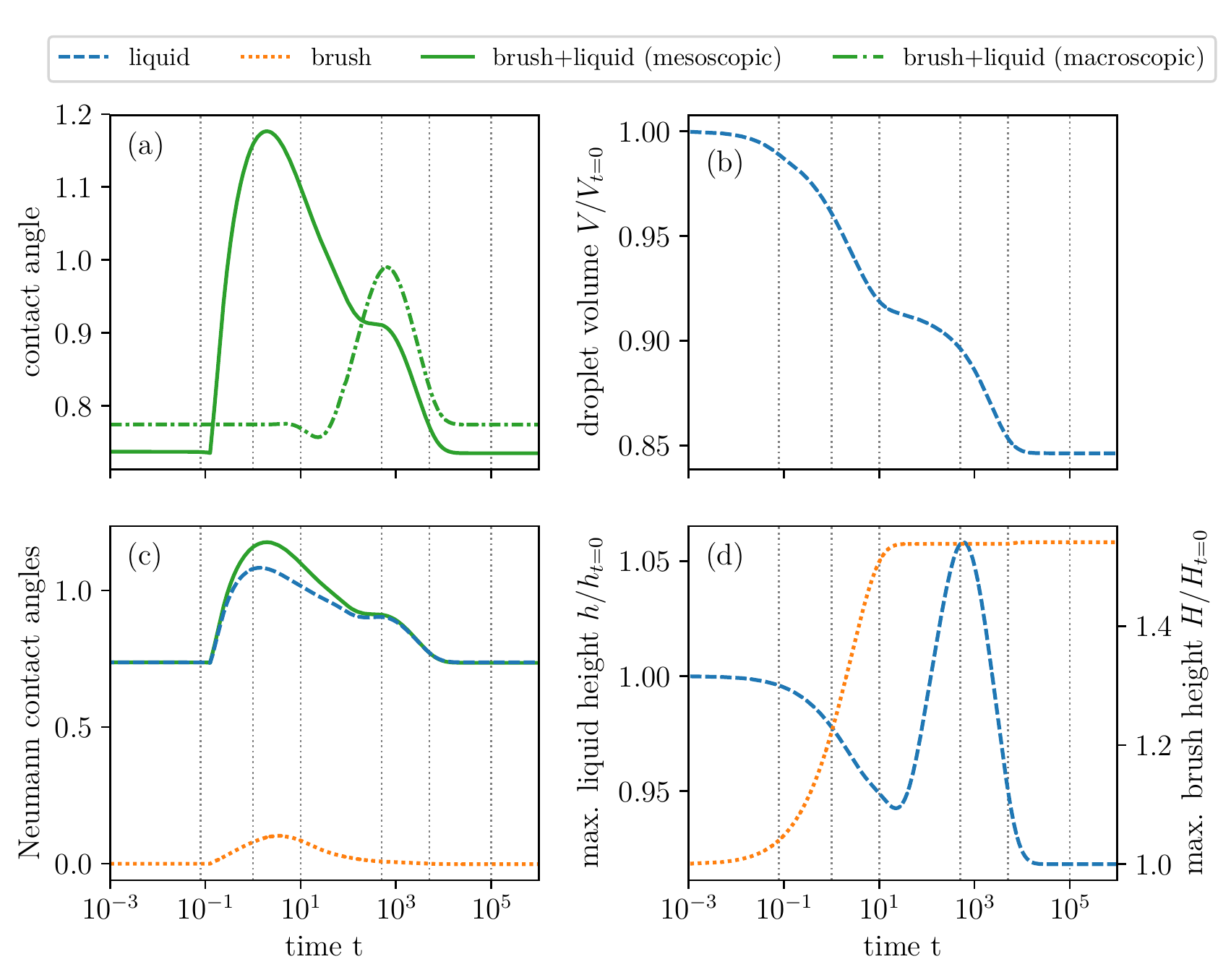}
  \caption{
For the spreading liquid drop on an initially dry polymer brush of Fig.~\ref{fig:buerste-spreading-snapshots} we give the time dependencies of (a) the macro- and mesoscopic contact angle, (b) the volume of the droplet, (c) the mesoscopic Neumann contact angles $(\theta_\zeta, \theta_h, \theta_{\zeta+h})$ and (d) the maximum heights of both liquid drop and brush. For a discussion of the used macroscopic and mesoscopic measures for the contact angle see main text. The times corresponding to the snapshots in Fig.~\ref{fig:buerste-spreading-snapshots} are indicated by thin vertical dotted lines. Note that the time axis is logarithmic.
}\label{fig:buerste-spreading-measures}
\end{figure}

The dimensionless dynamical model (\ref{eq:twofield-evolution-nondim}) can be employed to study the spreading of a droplet on an initially dry polymer brush. In the following we illustrate this by a direct numerical simulation. The results are given in Fig.~\ref{fig:buerste-spreading-snapshots} in the form of snapshots that illustrate the film height profile $h+H$ and the brush thickness profile $H$ as solid lines and the absorbed amount of liquid within the brush $\zeta$ by a yellow to orange shading of the brush layer. The insets show the respective contact line regions. The employed parameters are listed in the caption.
Fig.~\ref{fig:buerste-spreading-measures} presents the corresponding dependencies of useful quantities on time, including a comparison of the macroscopic and mesoscopic contact angles in Fig.~\ref{fig:buerste-spreading-measures}~(a).
In long-wave approximation, contact angles are given by the slope of the respective height profiles where they meet the substrate. In mesoscopic models they can be measured as slope at the inflection point of the liquid-gas interface. Note, however that the mesoscopic angle normally differs from the macroscopic one. The latter is obtained by extrapolation of a spherical cap shape from the drop maximum to the substrate. These quantities are well defined for equilibrium drops on rigid smooth substrates (see e.g.~\cite{TMTT2013jcp}).

However, the situation is trickier for the present situation where contact angles shall be measured in a dynamic situation where the substrate is not flat and rigid. We define mesoscopic and macroscopic angles in close analogy to the simple described case, however, one has to be careful when giving an interpretation.

We measure the three (mesoscopic) Neumann angles at the same position $x_c$ where the profile of the liquid layer has its inflection point, i.e., where 
\begin{equation}
    h'(x_c) = \max\,h'(x),
\end{equation}
The three angles in long-wave approximation are then 
\begin{equation}
    \theta_h = \frac{\partial h}{\partial x}\bigg|_{x=x_c},~~~\theta_\zeta = \frac{\partial \zeta}{\partial x}\bigg|_{x=x_c},~~~\theta_{\zeta+h} = \frac{\partial (\zeta+h)}{\partial x}\bigg|_{x=x_c}.
\end{equation}

In contrast, the macroscopic contact angle is defined as the slope of a parabolic fit $f(x)$ to the apex region of $h(x)+\zeta(x)$ at the intersection with a macroscopic approximate of the brush surface line:
\begin{equation}
    \theta_\mathrm{macro} = f'(x_s)~~~\text{with}~~~f(x_s) = \min (h(x)+\zeta(x)).
\end{equation}

Further, Fig.~\ref{fig:buerste-spreading-measures}~(b) shows the time evolution of the normalized droplet volume $V=\int \mathrm{d}x\,h$, Fig.~\ref{fig:buerste-spreading-measures}~(c) the three Neumann contact angles and Fig.~\ref{fig:buerste-spreading-measures}~(d) the normalized maximal heights $h_{\max}(t)=\max h(x,t)$ and $H_{\max}(t)=\max H(x, t)$. The logarithmic time-scale reveals that there are different phases of the spreading dynamics, indicating the interplay of different physical effects. In the following, we study these by means of Figs.~\ref{fig:buerste-spreading-snapshots} and~\ref{fig:buerste-spreading-measures}.

As initial condition ($t=0$), we use the equilibrium solution of a liquid drop on a rigid reference substrate, that we obtain from a simulation with vanishing transfer rate $M=0$. We assume the substrate to be nearly dry, setting $\zeta(x, t=0) = 0.1$. We then switch on the substrate dynamics $M>0$ and observe the dynamics until the simulation reaches the new equilibrium state.

Until $t\approx 0.2$, the initial conditions relax, i.e., the precursor layer and the brush absorption equilibrate, resulting in a rather small homogeneous swelling of the brush (cf.\ Fig.~\ref{fig:buerste-spreading-snapshots}~(a)).
Till $t\approx 2.0$, underneath the drop, liquid is absorbed into the brush that swells in consequence. Naturally, the droplet volume decreases and the contact line recedes slightly. Simultaneously, the mesoscopic contact angle and the other Neumann angles all increase, while the macroscopic contact angle remains constant as the dynamics mainly occurs in the contact line region (cf.\ Fig.~\ref{fig:buerste-spreading-snapshots}~(b)).
At $t\approx 10$, the swelling of the brush underneath the drop is nearly equilibrated, as the brush humidity diminishes the transfer rate. However, diffusion in the brush transports liquid away from the drop, where liquid is replenished by absorption. The mesoscopic angles start to decrease again, while the macroscopic one is still unchanged (cf.\ Fig.~\ref{fig:buerste-spreading-snapshots}~(c)).
Subsequently, the brush starts to swell in the region away from the droplet, leading to a further decrease of the mesoscopic contact angles. Then also the macroscopic angle reacts, first, it shortly decreases before it increases due to a steepening of the overall drop shape. The process peaks at $t\approx 10^3$ (cf.\ Fig.~\ref{fig:buerste-spreading-snapshots}~(d)).
Lastly, diffusion in the brush continues until the whole brush has equilibrated. In parallel, the drop spreads again, leading to a decrease of all contact angles. (cf.\ Figs.~\ref{fig:buerste-spreading-snapshots}~(e) and~(f)).
Beyond $t\approx 10^4$, the new equilibrium state is reached. As expected, the contact angles are the same as the equilibrium angles of the initial dry brush. This is due to our simplifying assumption that we explicitly exclude a dependency of wettability and brush-liquid interface tension on the brush state.

\begin{figure}[thb]
    \centering
    \includegraphics[width=0.9\textwidth]{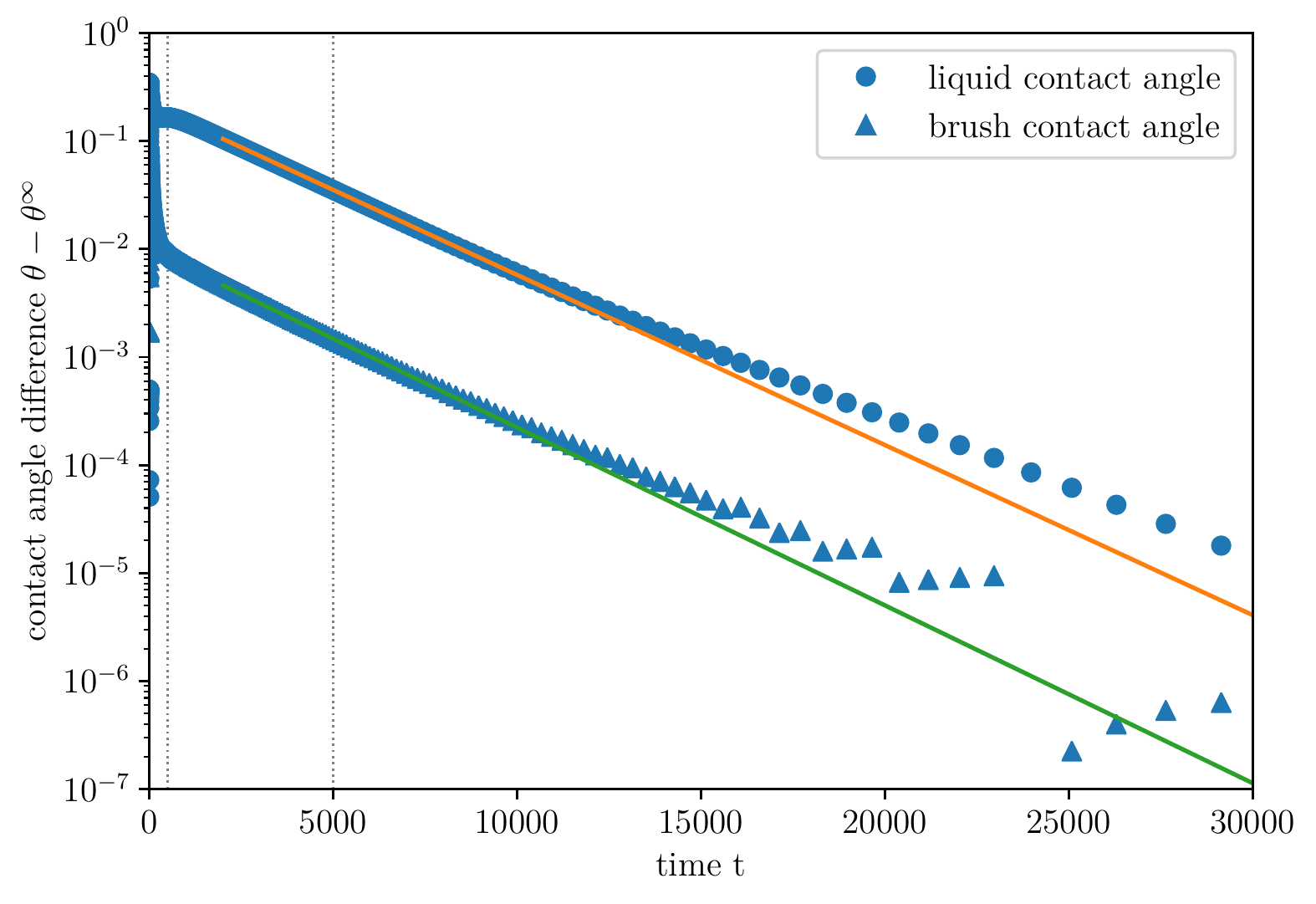}
    \caption{Log-normal plot of the mesoscopic contact angles $\theta_h$ (liquid) and $\theta_\zeta$ (brush) as a function of time showing their exponential approach of their equilibrium values $\theta_h^\infty$ and $\theta_\zeta^\infty$. The measured values (symbols) may be approximated by an exponential $f(t)=\lambda \exp(-t/\tau)$ as indicated by the solid lines. When the difference becomes very small, numerical precision is lost.}
    \label{fig:buerste-exp-decay}
\end{figure}

In~\cite{BBSV2018l} it is assumed that after placing a drop on an adaptive substrate, the interfacial tension relaxes exponentially towards a new equilibrium value, i.e., the contact angle at the end of the simulation should show the same behaviour. However, it is also mentioned that the observed change can be complex and depends on the particular considered process. Even though, for simplicity, our initial and final equilibrium contact angles are chosen to be identical, our simulations confirm both aspects. On the one hand, we find a rather complex dynamics of the relevant contact angles (see Fig.~\ref{fig:buerste-spreading-measures} and discussion above) that reflect the interplay of the coupled relaxation processes. On the other hand, the final equilibrium is approached via an exponential relaxation as evidenced in the log-normal plot in Fig.~\ref{fig:buerste-exp-decay}. There, we show the exponential decay of the mesoscopic contact angles $\theta_{h}$ and $\theta_\zeta$ towards their equilibrium values $\theta_h^\infty$ and $\theta_\zeta^\infty$ and compare the simulation results (symbols) to an exponential fit in the relaxation range $2\cdot10^3<t<3\cdot10^4$. For large times the difference becomes very small, and a high sensitivity towards finite grid effects and a precision loss in numerical subtraction become visible. Nevertheless, the relaxation of the contact angles, and thus, the respective interfacial tensions is well approximated by an exponential.

\section{Conclusion}\label{sec:conc}
We have presented a physico-chemical mesoscopic hydrodynamic thin-film model for the droplet spreading on an adaptive substrate. The model has been developed as a gradient dynamics on an underlying energy functional. After presenting the general framework we have considered the coupled spreading, absorption, diffusion and swelling dynamics that occurs when a liquid drop is placed on a dry polymer brush. There the underlying energy functional accounts for capillarity, wettability and brush energy. Finally we have employed the basic model to numerically simulate a droplet spreading on a swelling brush. The analysis has shown an intricate dynamics consisting of several phases with qualitatively different behaviour. It will be an interesting for future studies to investigate in detail how the behaviour changes when the ratios of the time scales for the various involved processes are varied.

The presented model is basic in some aspects and allows for several pathways of future improvements: (i) It has been assumed that there is no dynamic coupling between liquid motion in the brush layer and the liquid in the drop. In consequence, only the elements of the main diagonal of the mobility matrix of the conserved dynamics are nonzero. Treating the liquid flow within the brush layer in detail will be cumbersome but feasible (cf.~analogous case of coupled flow on and in a thin porous layer~\cite{ThGV2009pf}), but main features can be captured by a brush-state dependent effective slip at the brush-liquid interface. Similar considerations will apply for drop spreading on a porous layer (cf.~systems described in \cite{Gamb2014cocis}). (ii) For simplicity here both, the brush-liquid interface tension and the wetting energy, have been chosen to be independent of the brush state. This can be amended by respectively using a brush-liquid interface energy $\gamma_\mathrm{bl}(\zeta) (1+(\partial_x\zeta)^2)$ [Eq.~(\ref{eq:brush-en-cap})] and a $f_\mathrm{wet}(h, \zeta)$ that explicitly depend on absorption $\zeta$. Note, however, that such dependencies have to be introduced in a consistent manner as recently discussed for surfactant-covered drops/films~\cite{TSTJ2018l}. Note, that a brush-state dependent brush-liquid interface energy may result in additional Marangoni-type flows, and the appropriate mobilities have to be discussed (cf.~\cite{ThAP2016prf} for the case of surfactants). (iii) Another such improvement is to replace the long-wave approximation of the interface energies by the full expressions, i.e., using the \enquote{exact-curvature trick}  of~\cite{GaRa1988ces,SADF2007jfm,BoTH2018jfm}.
Of these amendments, the one concerning the mobilities, i.e.~(i), will most likely be less important than the ones improving on the energies, i.e.~(ii) and~(iii). For other thin-film models this is discussed in~\cite{Thie2018csa}. 

More generally, the modelling approach can be applied to more complex situations by expanding the framework we have presented in section~\ref{sec:model}. Evaporation of liquid from the drop and the brush as well as condensation into the brush may be incorporated via further nonconserved terms in Eqs.~(\ref{eq:twofield-evolution-one}) and~(\ref{eq:twofield-evolution-two}) that do not balance each other. This would capture the case of phase-transition controlled evaporation/condensation as considered in~\cite{LyGP2002pre, Thie2010jpcm,EnTh2019el}, also see discussion on pgs.~404--405 of~\cite{Thie2014acis}. Further, changing the brush energy should allow one to consider a liquid layer on a brush in the case where the liquid is not fully miscible with the brush.  Studies employing MD simulations \cite{MeSB2019m} have shown that then phase transitions between a partially wetting drop on a dry brush, a wetting liquid layer on a dry brush and a fully mixed homogeneous brush state occur. Also an adaptive substrate interacting with a liquid mixture is of high interest, see e.g., recent works on the co-nonsolvency transition~\cite{BKTW2017l,Somm2017m,YBUF2019m}. To model drops of mixtures on brushes one would need to combine the approach presented here with gradient dynamics models for films of mixtures~\cite{ThTL2013prl}. In general, it should be possible to account for most of the adaptive substrates discussed in the introduction of~\cite{BBSV2018l} with gradient dynamics models extending the one we have presented here.

\section*{Acknowledgement} We acknowledge support by 
the Deutsche Forschungsgemeinschaft (DFG, Grant No.\ TH781/12 within SPP~2171).
The authors also express their gratitude for all discussions during the workshop \enquote{Challenges in nanoscale physics of wetting phenomena} organised by S.~Afkhami, T.~Gambaryan-Roisman and L.~Pismen; and hosted by the Max Planck Institute for the Physics of Complex Systems in Dresden, Germany, from 26--30 August 2019. In particular, we thank G.~Auernhammer and S.~Schubotz for sharing their unpublished results on droplet spreading on polymer brushes.

\bibliography{ThHa2019}
\end{document}